\documentstyle[aps,prl,amssymb,graphicx,preprint]{revtex}

\begin{document}

\title{X-Ray Scattering Evidence for Macroscopic Strong Pinning Centers in the Sliding CDW state of NbSe$_3$.}

\author{D. Rideau $^{1,2}$, P. Monceau $^1$, R. Currat $^2$,
H. Requardt $^{3,6}$, F. Nad $^{1,4}$, J.E. Lorenzo $^{5,6}$, S. Brazovskii $^7$, C. Detlefs $^6$ and G. Gr\"uŸbel $^6$ }
\address{$^1$ Centre de Recherches sur les Tr\`es Basses TempŽ\'eratures, CNRS, B.P. 166, 38042 Grenoble, France.\\ 
 $^2$ Institut Laue-Langevin, B.P. 156, 38042 Grenoble, France. \\
$^3$ MPI fŸ\"ur Metallforschung, Heisenbergstr. 1, 70569 Stuttgart, Germany.\\
$^4$ Institute of Radio-Engineering and Electronics, 103907 Moscow, Russia.\\
 $^5$ Laboratoire de Cristallographie, CNRS, B.P. 166, 38042 Grenoble France.\\
 $^6$ ESRF, B.P. 220, 38043 Grenoble, France.\\
$^7$ LPTMS, B‰t. 100, 91406 Orsay, France.\\}

\date{\today} 

\maketitle 

\begin{abstract}
Using high-resolution X-ray scattering techniques, we measure the variation, $q(x)$, of the position in reciprocal space of the CDW satellite, in the sliding state, along the length of NbSe$_3$ whiskers. We show that structural defects and intentionally X-ray radiation-damaged regions increase locally the CDW pinning force, and induce CDW phase distortions which are consistent with those observed near contacts. Using the semi-microscopic model in Ref. \cite{R16} describing the normal $\leftrightarrow$ condensed carrier conversion, with spatially varying parameters, we account for the experimental spatial dependence of the CDW phase gradient near both types of defects.

\vspace*{2mm} 
\end{abstract}
 
\vspace{2mm} 
\narrowtext

 In strongly anisotropic charge density wave (CDW) systems a phase transition of mixed structural and electronic character occurs at the Peierls temperature. The CDW ground state is characterized by a periodic modulation of the crystal lattice, accompanied by a periodic modulation of the electron density of same wavevector $Q_0=2k_F$, where $k_F$ is the Fermi wave number \cite{R1,R3,Poug}.

In certain CDW materials, such as NbSe$_3$, in which the wavelength of the CDW is incommensurate with the underlying lattice, sliding of the CDW as a collective motion is made possible by the application of an electrical current exceeding a threshold value $I_T$. This collective charge transport is associated with remarkable electrical properties such as non-ohmic conductivity and narrow band noise (NBN) \cite{elec}.

Phase slippage is the process by which conversion between free electron and CDW condensate occurs near the electrical contacts to a CDW conductor \cite{R4}. It proceeds via nucleation and growth of $2 \pi$ phase dislocations loops (DL) across the sample cross section \cite{R5}. Associated with the DL nucleation and growth process, the sliding-CDW phase is slightly distorted. The phase gradient is observable by means of X-ray diffraction as a longitudinal shift, $q \varpropto \partial \phi/ \partial x$, of the CDW satellite peak position in reciprocal space, $Q=Q_0+q$ \cite{R6,RR}.
 The CDW phase gradient has also been monitored by means of laser probe \cite{R7}, electromodulated transmission \cite{R8},  and conductivity measurements on multicontacted samples \cite{R9}. Even though all experimental data show {\it qualitative} agreement (the phase gradient exhibits maxima in the near contact regions and vanishes in the central part of the sample) there remains some controversies about the nature of the CDW-defect interaction \cite{last,Bal,Poug1}. While the above experiments reported phase distortions due to conversion {\em near the electrodes}, phase slippage is potentially needed near all defects of the {\em strong pinning} type.

 In this paper, we show that a localised defect may act as a pseudo-electrode. Using high resolution X-ray diffraction, we monitored the shift, $q(x)$, of the NbSe$_3$ sliding-CDW satellite as a function of beam position, $x$, along two samples which exhibit, respectively, one and two localised structural defects between current electrodes. In addition to the near contact region, where the sliding-CDW phase is substantially distorted, we observe a similar phase distortion in the vicinity of the defects. A further experiment was performed by exposing part of sample '2' to the X-ray beam, over a long period of time, in order to damage it locally. In each case, the observed phase gradients are compared to numerical simulations based on the semi-microscopic model by Brazovskii {\it et al.} \cite{R16}, describing the normal $\leftrightarrow$ condensed carrier conversion. We give an estimate of the fraction of the CDW current which is converted, at the defect position.

 NbSe$_3$ has a chain-like structure, with chains parallel to the monoclinic $b$-direction. It undergoes two independent Peierls transitions at 145 K and 59 K with modulation wave vectors $(0,Q_0,0)$; $Q_0$=0.241 b$^*$ and $(0.5,Q_1,0.5)$; $Q_1$=0.259 b$^*$, respectively. 
All measurements were carried out on the $(0,1+Q_0,0)$ CDW satellite on the diffractometer TROIKA I (ID 10A) at ESRF 
(Grenoble/France) using an incident wavelength of 1.127 \.A ($E = 11$ keV). The sample orientation was such that the (a$^*$+c$^*$, b$^*$)-plane coincided with the horizontal scattering plane. In order to obtain a high spatial resolution, the beam width (from 10 to 30 ${\rm \mu}$m) was controlled by a slit, placed before the sample.
The samples, of typical cross-sections 10$*$2 $\mu $m$^2$, were mounted on sapphire substrates of 100 ${\rm \mu}$m thickness to provide homogeneous sample cooling together with suitable beam transmission (50 \%). Several types of electrical contacts were used. Two 2 ${\rm \mu}$m-thick gold electrodes, of 1 mm width, have been evaporated on Sample '1', leaving an uncovered section of 4.25 mm between electrodes. On sample '2', one 1 mm-wide electrode and one 15 ${\rm \mu}$m-wide electrode have been evaporated, leaving an uncovered section of 3 mm between electrodes.

The samples are selected for their high crystallographic quality. We scanned the (020) Bragg peak along the b* direction (($\theta,2 \theta$) scans) and along the a*+c* direction (rocking scans), for several beam positions ($50$ to $100$ $ {\rm \mu}$m intervals). We tested a large number of samples from different origins  \cite{T}, and we found that localised ($\approx 100$ ${\rm \mu}$m-wide) growth defects appear very frequently in this material \cite{characterisation}. 

 Fig. 1 shows the shift, $q(x)=Q_I(x)-Q_0$, for positive and negative current polarities; I=$\pm 4.6$ mA ($I$/$I_T$=2.13), as a function of beam position, $x$, along sample '1'. This sample exhibits, between electrodes, one localised defect at the position $x_d \approx -.15$. The sliding-CDW satellite shift changes sign abruptly at this position, with maxima on either sides of the defect position, as well as at the electrical contacts. {\em In contrast}, for defect-free samples, the phase gradient is large near the electrodes, where the conversion between normal carriers and CDW condensate occurs, and vanishes in the central section \cite{R16,R6}.
 
Fig. 2 shows the shift, $q(x)$, for positive and negative current polarities, I=3.5 mA ($I$/$I_T$=3), along sample '2', which exhibits, two localised defects at $x_1 \approx -.5$ and $x_2 \approx .75$. Again the shift changes value abruptly at these positions. Note incidentally that the CDW distortions extend over an appreciable distance ($\lambda_1 \approx 100$ ${\rm \mu}$m) beyond the contact boundary, regardless of the electrode width (15 ${\rm \mu}$m or $1$ mm). This behaviour has been confirmed on defect-free samples \cite{unpub}.
 
 A further experiment was performed, by exposing sample '2' to a 30 ${\rm \mu}$m-wide X-ray beam, over a long period of time (4 hours), in order to locally create a damaged region. After the irradiation, the CDW satellite position changes value abruptly at the irradiated position ($x_d=-1.27$; the arrow position in Fig. 3).  The open symbols in Fig. 3  show the shifts, $q(x)$, for positive and negative current polarities, as a function of beam position, $x$, before the irradiation. The shift, $q(x)$, follows a typical  monotonic decay in the near-contact region. The same measurements repeated after irradiation show a drastic change of the spatial dependence of the shift over a distance of approximately 200 ${\rm \mu}$m on either side of the irradiated position. The damage is also observed as a local broadening of the (020) Bragg reflection; this broadening is detected over a sample length of $20$ $\pm 5 {\rm \mu}$m around $x_d$, which is consistent with the width of the irradiating beam. No significant longitudinal splitting of the CDW satellite was observed, near the defect, indicating that the cross section is homogeneously affected by the irradiation.

The semi-microscopic model by Brazovskii {\it et al.} \cite{R16}, describing the normal $\leftrightarrow$ condensed carrier conversion by nucleation and growth of dislocation loops in a highly rigid CDW electronic crystal gives a coherent fit of the spatial dependence of the CDW phase gradient, for homogeneous samples \cite{R16,R6,R17}. This model is based on the assumption of a local equilibrium between the electrochemical potentials of the phase dislocations, $U$, and of the free carriers, $\mu_n$: 
\begin{equation}
\frac{\partial \eta }{\partial x}=F_r(J_{c})-\frac{e(J_n-J_T)}{
 \sigma _{n}}  \label{eta'}
\end{equation}
where $J_{n}$, $J_c$ and $J_{tot}=J_n+J_c$ are, respectively, the normal carrier, CDW and total current densities. $J_T$ is the current density at threshold, and $\sigma_n$, the normal carrier conductivity. For high current values ($J_{tot} \geqq 2J_T$), the friction force, $F_r(J_c)$, can be approximated as $F_r(J_c) \approx eJ_c / \sigma _{c}$, where $\sigma_c$ is the high-field CDW conductivity. $\eta$ measures the deviation of the normal carrier concentration from its $I=0$ value:
\begin{equation}
\eta \equiv K^{eff}q=\mu _{n}-U 
\end{equation}

Eq. (1) is equivalent to the one-dimensional FLR equation \cite{FLR}, with an additional contribution ($\varpropto \beta_e $) to the effective elastic constant per unit length, $K^{eff}=gS/2\pi N_{F}^{i}$, characterizing the screening of the CDW distortions by the carriers from uncondensed bands (in NbSe$_3$, at these intermediate temperatures (90 K), $\beta_e \approx 0.65$); $g^{-1}=(\frac{\rho_s}{ \rho _{c}}+\beta _{e})$, where $S$ is a chain cross-section, $\rho_c$ the CDW density, $\rho_s=1-\rho_c$ and $N_F^i$ is the density of states at the Fermi level in the undistorded metallic state. 

The gradient of the CDW current density is controlled by the conversion rate, $R$, between normal carriers and the CDW condensate:
\begin{equation}
\partial J_{c}/\partial x=2(eg)R(\eta,J_{c})+\nu_{\rm inj} 
\label{Jc}
\end{equation}
with $\nu_{\rm inj}=J_{tot}(\delta_a(x)-\delta_{-a}(x))$ the injection-extraction term (for contacts at $x=\pm a$), where $\delta_a(x)$ is a Dirac function centered at a.
As noticed previously \cite{R9}, the earlier function, $R \varpropto e^{-V_a/|q|}$, describing a stress-initiated nucleation of DL, proposed by Ramakrishna {\it et al.} \cite{rama}, leads to an unsatisfactory relation between the CDW current, and the CDW phase gradient. This function does not, either, account for our experimental data. As discussed in detail in \cite{R16,R6,R17}, eqs. (1) and (3), with the empirical function $R(\eta)=r_0 \eta$, ($r_0=g\tau_{{\rm cnv}}^{-1}/2 \pi N_F^iS$ is treated as an adjustable parameter), gives a coherent fit for the homogeneous situation. Here $\tau _{{\rm cnv}}$ is the lifetime of an excess carrier with respect
to its conversion to the condensate, that is the mean free carrier lifetime before
absorption by a DL. 
The parameters relevant to the {\em homogeneous} case and its solution will be
labeled by the subscript $0$: 
\begin{equation}
\eta _{0}(x) \quad \varpropto \quad \frac{\sinh
x/\lambda _{0}}{\cosh a/\lambda _{0}}  \label{eta}
\end{equation}
where $\lambda _{0}=\sqrt{\frac{S K^{eff}}{e^{2}}\sigma _{0}^{\ast }\tau _{{\rm %
cnv}}}$, typically a few hundred ${\rm \mu}$m, is the characteristic length
scale of the phase slip distribution, and $\sigma _{0}^{\ast }=(1/\sigma _c^0+1/\sigma _n^0)^{-1}$ is the high-field conductivity. Far from contacts ($|x-a| \geqq 3 \lambda_0$), where the equilibrium condition $\mu_n=U$ is achieved ({\it i.e. $\eta \varpropto q=0$}), $\frac{\partial \eta }{\partial x}=0$. 

For {\em inhomogeneous} samples, $\frac{\partial \eta }{\partial x}$ can also locally vanish even though $\eta \neq 0$.
This  case can be studied with the help of the
same equations (1) and (3), but with {\em spatially varying parameters}; $\sigma_c(x)=\sigma_c^0-\Delta \sigma_c f(x-x_d)$ and $r_0$ become dependent on $x$. $f(x)$ is a fast decaying function, introduced in order to take account of the finite size of the defect. The friction force becomes 
\begin{equation}
F_r=\frac{eJ_c}{\sigma_c^0-\Delta \sigma_c f(x-x_d)} \approx \frac{eJ_c}{\sigma_c^0} +eJ_c^0 \frac{\Delta \sigma_c f(x-x_d)}{(\sigma_c^0)^2}
\end{equation}
and the phase slip rate becomes
\begin{equation}
 R(\eta,x)=r_0\eta+\Delta r_0\eta f (x-x_d)
\end{equation}
The pinning
increase under irradiation is a well known effect \cite{last,irrad}, but the effect of irradiation upon the
conversion rate, $R$, is more subtle: on one hand,  $R$ is likely to increase together with the
number of nucleation centers but at the same time it may decrease because of the enhanced pinning force against the lateral motion of the
dislocation lines. In what follows, we consider only $F_r$,
{\it i.e.} $\sigma _{c}$ to be $x$ dependent ($\Delta r_0=0$).  
According to the above hypotheses, and
equating $f(x)$ to unity when $-\zeta/2<x<\zeta/2$ and to zero elsewhere ($\zeta$ is the size of the defect), eqs. (1) and (3) reduce to: 
 \begin{equation}
\frac{\partial^{2} \eta}{\partial x ^{2}}=\frac{2e}{\sigma _{0}^{\ast }}r_0\eta + eJ_c^0 \frac{\Delta \sigma_c}{(\sigma_c^0)^2}[ \delta_{x_d-\frac{\zeta}{2}}- \delta_{x_d+\frac{\zeta}{2}}]+e\frac{\nu_{\rm inj}}{\sigma_c^0}
\end{equation}
with the boundary conditions $J_{tot}=0$ at $|x| = a $.
Here the delta function provides a partial conversion of the CDW current at the defect position. This term has indeed the same mathematical form as $\nu_{\rm inj}$. In the sliding  state, a fraction, ${\Delta \sigma_c}/{\sigma_c^0}$, of the CDW current, $J_c^0(x)$, is transformed into normal carriers (and recondensed immediately after the defect position). The defect can be considered as a pseudo contact, at the boundary of which, depending on the value of $\Delta \sigma_c$, the conversion is total or partial. For $\Delta \sigma_c/{\sigma_c^0} \approx 1$, all the CDW current is transformed, and the shift changes sign as seen in Fig. 1, whereas low values of $\Delta \sigma_c/{\sigma_c^0}$, lead to the behaviour shown in Figs. 2 and 3.

 In practice, in the numerical simulation, a more realistic gaussian function, centered at the defect position (width=$\zeta$), has been used for $f$.  Beneath the contacts we used the same general equations, setting the last term in eq. (1) equal to zero. Eq. (7) then becomes a diffusion equation for the excess of free carriers, $\eta$. Its solution
 matches the inner solution at $x= \pm a$ and decreases exponentially, beneath contacts, with a characteristic length $\lambda_1=\sqrt{(\sigma_c^0/\sigma_0^{\ast })} \lambda_0 \approx \lambda_0/2$.

 The best fit to the data in Fig. 1 is obtained with $\lambda_0=330$ ${\rm \mu}$m, $\lambda_1=100$ ${\rm \mu}$m, $\zeta=72$ ${\rm \mu}$m and $\Delta \sigma_c/{\sigma_c^0}=0.47$ \cite{assym}. This last value implies that approximately half of the CDW current is converted into normal carriers at the defect position and recondensed just after. This significant phase slippage provides for the spectacular jump in $q$ in the central part of the sample.  The slight discrepancies between our experimental data and the calculated CDW phase distortions may come from the fact that $\sigma_c^0$ is not perfectly constant along the sample or from our simplifying assumption $\Delta r_0=0$ \cite{attempt}. 

The best fit to the data in Fig. 3 is obtained with $\lambda_0=245$ ${\rm \mu}$m, $\lambda_1=100$ ${\rm \mu}$m, before irradiation, and $\lambda_0=196$ ${\rm \mu}$m, $\lambda_1=75$ ${\rm \mu}$m, $\zeta=38$ ${\rm \mu}$m (which is consistent with the size of the irradiating beam) and $\Delta \sigma_c/{\sigma_c^0}=0.45$, after irradiation. Again, the defect provides for a strong conversion, but in contrast to Fig.1, the defect is located close to a contact boundary where the CDW current has not yet reached its maximum value $|J_c^0|_{{\rm max}}$, where, typically at 90 K,  $|J_c^0|_{{\rm max}} \approx \frac{1}{3} J_{tot}$, and hence the shift, $q$, does not change sign.

In conclusion, we have observed the influence of localised defects on the longitudinal variation, $q_I(x)$, of the CDW satellite position at fixed DC current, I, as a function of beam position, $x$, between electrical contacts, in NbSe$_3$. We have shown, that the sliding-CDW phase distortion, due to conversion between normal carriers and CDW condensate, is not uniquely a near-contact effect but also occurs  in the vicinity of structural defects or regions intentionally damaged by X-ray radiation.  Using the semi-microscopic model by Brazovskii {\it et al.}, with an enhanced pinning force at the defect position, we have given a coherent description of the spatial dependence of the CDW phase gradient near both types of defects. The phase gradient, which is very similar to what is observed at contacts (for several types of electrodes), reflects the fact that a fraction of the CDW current is transformed into normal current at the defect position. 

The authors are indebted to L. Ortega and G. Patrat (Cristallographie/CNRS) for making possible extensive sample quality tests.

%

\begin{figure}
\begin{center} 
\includegraphics[scale=1]{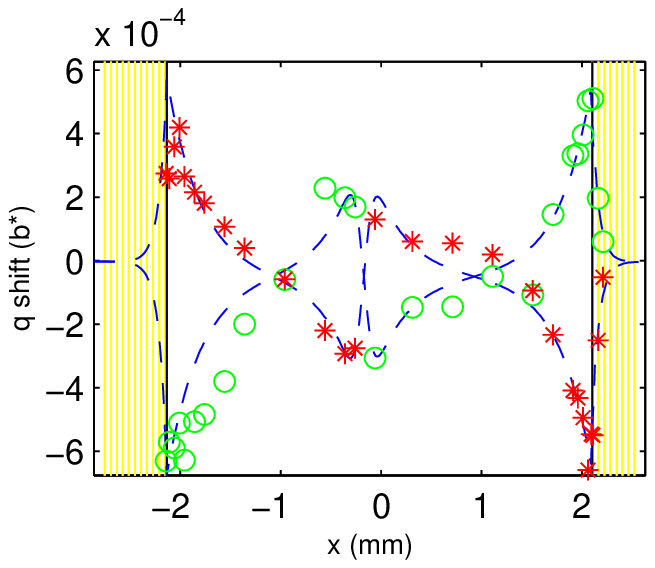} 
\caption{   Shift $q(x)=Q_I(x)-Q_0$ of the CDW satellite peak position as a function of beam position along sample '1' for positive (*) and negative (o) polarities; I=$\pm 4.6$ mA (I/$I_T$=2.13).  The vertical lines show the boundaries of the gold-covered contacts (dashed area); dashed lines are the best fit solutions to eq. (8);  beam width: 30 ${\rm \mu}$m; $NbSe_3$; T=90 K.  
}\label{figurename}\end{center}\end{figure}

\begin{figure}[b]

\begin{center}

\includegraphics[scale=1]{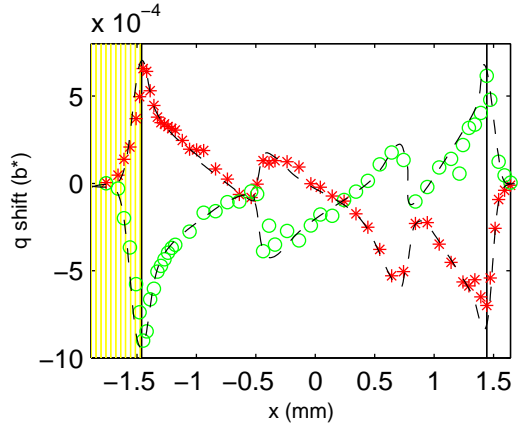} 
\caption{ 
Shift, $q(x)$, along sample '2' before irradiation for positive (*) and negative (o) polarities; I=$\pm 3.5$ mA (I/$I_T$=3).  The vertical lines show the boundary of the 1 mm-wide gold-covered lhs contact (dashed area) and the 15 $\mu m$-wide rhs contact. Dashed lines are guides for the eye; beam width: 30 ${\rm \mu}$m; NbSe$_3$; T=90 K.}\label{figurename2}\end{center}\end{figure}

\begin{figure}
\begin{center}\leavevmode
\includegraphics[scale=1]{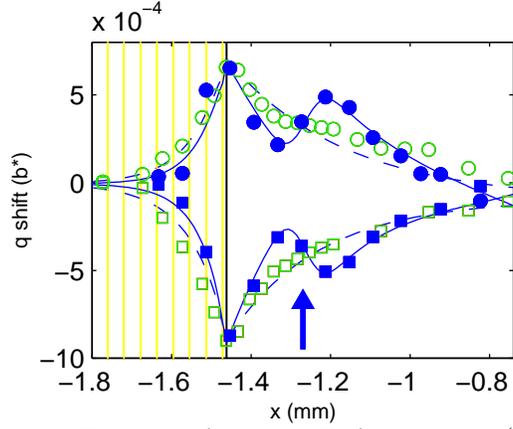}
\caption{  Shift, $q(x)$, along the lhs part of sample '2' for positive (o) and negative ({\tiny $\square$}) polarities; I=$\pm 3.5$ mA (I/$I_T$=3). Full symbols show the shift q($\pm I$) after a local irradiation at $x_d$=-1.27 (arrow). The vertical line shows the boundary of the lhs gold-covered contact (the dashed area). Dashed and full lines are the best fit solutions to eq. (8);  NbSe$_3$; T=90 K.}\label{figurename3}\end{center}\end{figure}   


\begin{thebibliography}{9}
\bibitem{characterisation} The mosaicity of samples '1' and '2' is found to be below 0.013$^o$, except around the defect positions. Defects appear to be related to clusters of impurities, which provide local misorientations of the  b$^*$-axis.
\bibitem{attempt} Attempts at varying locally $r_0$, rather than $\sigma_c$, yield only a change of slope for $q(x)$ and fail to produce a non-monotonic behaviour. However, the assumption of $r_0$ to be x-independent is not exact. The slight difference in $\lambda \varpropto \sqrt{r_0}$ before and after irradiation shows indeed that $r_0$ is slightly sensitive to irradiation defects.
\end{thebibliography}
\end{document}